\documentclass{aa}

\begin{document}

\title{The mass of the neutron star in Vela X-1 and
tidally induced non-radial oscillations in GP Vel}

\titlerunning{Neutron star mass in Vela X-1}

\author{H. Quaintrell\inst{1} \and  A.J. Norton\inst{1} \and 
T.D.C. Ash\inst{1,2} \and P. Roche\inst{1,3,4} \and B. Willems\inst{1}
\and T.R. Bedding\inst{5} \and \\ I.K. Baldry\inst{5,6} \and 
R.P. Fender\inst{7}}

\authorrunning{Quaintrell, Norton, Ash et al}

\offprints{A.J. Norton, a.j.norton@open.ac.uk}

\institute{Department of Physics and Astronomy, The Open University, 
        Walton Hall, Milton Keynes MK7 6AA, U.K.
\and
Kildrummy Technologies Ltd, 1 Mill Lane, Lerwick, Shetland, ZE1 0AZ, U.K.
\and
Earth and Space Sciences, School of Applied Sciences, University of Glamorgan,
        Pontypridd CF37 1DL, U.K.
\and
Department of Physics and Astronomy, University of Leicester, 
        Leicester, LE1 7RH, U.K.
\and
School of Physics, University of Sydney, New South Wales, 2006, Australia
\and
Department of Physics and Astronomy, The Johns Hopkins University,
        3400 North Charles Street, Baltimore, MD 21218-2686, U.S.A.
\and
Sterrenkundig Instituut `Anton Pannekoek', Kruislaan 403, 1098 SJ Amsterdam,
The Netherlands
}
  
\date{Accepted ???
      Received ???}

\abstract{ We report new radial velocity observations of GP~Vel / HD~77581, the
optical companion to the eclipsing X-ray pulsar Vela~X-1. Using data spanning
more than two complete orbits of the system, we detect evidence for tidally
induced non-radial oscillations on the surface of GP~Vel, apparent as peaks in
the power spectrum of the residuals to the radial velocity curve fit. By
removing the effect of these oscillations (to first order) and binning the
radial velocities, we have determined the semi-amplitude of the radial velocity
curve of GP~Vel to be $K_{\rm o} = 22.6 \pm 1.5$~km~s$^{-1}$.  Given the
accurately measured semi-amplitude of the pulsar's orbit, the mass ratio of the
system is $0.081 \pm 0.005$. We are able to set upper and lower limits on the
masses of the component stars as follows.  Assuming GP~Vel fills its Roche lobe
then the inclination angle of the system, $i$, is $70.1^{\circ} \pm
2.6^{\circ}$. In this case we obtain the masses of the two stars as $M_{\rm x}
= 2.27 \pm 0.17$~M$_{\odot}$ for the neutron star and $M_{\rm o} = 27.9 \pm
1.3$~M$_{\odot}$ for GP~Vel.  Conversely, assuming the inclination angle is
$i=90^{\circ}$, the ratio of the radius of GP~Vel to the radius of its Roche
lobe is $\beta = 0.89 \pm 0.03$ and the masses of the two stars are $M_{\rm x}
= 1.88 \pm 0.13$~M$_{\odot}$ and $M_{\rm o} = 23.1 \pm 0.2$~M$_{\odot}$. A
range of solutions between these two sets of limits is also possible,
corresponding to other combinations of $i$ and $\beta$. In addition, we note
that if the zero phase of the radial velocity curve is allowed as a free
parameter, rather than constrained by the X-ray ephemeris, a significantly
improved fit is obtained with an amplitude of $21.2 \pm 0.7$~km~s$^{-1}$ and a
phase shift of $0.033 \pm 0.007$ in true anomaly. The apparent shift in the
zero phase of the radial velocity curve may indicate the presence of an
additional radial velocity component at the orbital period.  This may be
another manifestation of the tidally induced non-radial oscillations and
provides an additional source of uncertainty in the determination of the 
orbital radial velocity amplitude.

 \keywords{
 binaries: close -- stars: neutron -- 
 stars: individual: Vela X-1 -- 
 stars: individual: GP~Vel 
 stars: individual: HD~77581 -- stars: fundamental parameters
 }}

\maketitle

\section{Introduction}

Binary systems containing an X-ray pulsar are very important
astrophysically as they can offer a direct measurement of the neutron star
mass. The mass ratio of the system, $q$, is simply given by the ratio of the
radial velocity semi-amplitudes for each component:

\begin{equation} 
q = \frac{M_{\rm x}}{M_{\rm o}}=\frac{K_{\rm o}}{K_{\rm x}}
\end{equation}

where $M_{\rm x}$ is the neutron star mass, $M_{\rm o}$ is the mass of the
optical component, $K_{\rm o}$ is the semi-amplitude of the radial velocity of
the optical component, and $K_{\rm x}$ is the semi-amplitude of the radial
velocity of the neutron star. In addition, for an elliptical orbit, it can be
shown that, 

\begin{equation} 
M_{\rm o} = \frac{K_{\rm x}^{3}P \left( 1-e^2 \right)^{\frac32}}
{2\pi G \sin^{3} i} \left( 1+q \right)^{2}
\end{equation}

and similarly,

\begin{equation}
M_{\rm x} = \frac{K_{\rm o}^{3}P \left( 1-e^2 \right)^{\frac32}}
{2\pi G \sin^{3} i} \left( 1+\frac{1}{q}\right)^{2}
\end{equation} 

where $i$ is the inclination of the orbital plane to the line of sight, 
$e$ is the eccentricity and $P$ is the period of the orbit.

We therefore have a means of calculating the mass of the neutron star if 
the orbits of the two components and the inclination of the system are known.
Such a combination is possible in an eclipsing X-ray binary system, in 
which the neutron star is a pulsar. X-ray pulse timing delays around the 
neutron star orbit yield the value of $K_{\rm x}$, and conventional radial 
velocity measurements from optical spectra yield $K_{\rm o}$. 

A value for $i$ can be obtained from the following approximations:

\begin{equation} 
\sin i \approx \frac{\left(1 - 
\beta^{2}\left(\frac{R_{L}}{a^{\prime}}\right)^{2}\right)^{\frac{1}{2}}}
 {\cos \theta_{e}} 
\end{equation} 

\begin{equation} 
\frac{R_{L}}{a^{\prime}} \approx A + B \log q + C \log^{2} q 
\end{equation} 

where  $R_{L}$ is the radius of the optical
companion's Roche lobe, $\beta = R_{\rm o}/R_{\rm L}$ is the ratio of the
radius of the optical companion to that of its Roche lobe, $a^\prime$ 
is the separation between the centres of masses of the two components, and 
$\theta_{e}$ is the eclipse half-angle. $A$, $B$, and $C$ have been 
determined by Rappaport \& Joss (1983) to be: 

\begin{equation} 
A \approx 0.398 - 0.026 \Omega^{2} + 0.004 \Omega^{3} 
\end{equation} 

\begin{equation} 
B \approx -0.264 + 0.052 \Omega^{2} - 0.015 \Omega^{3} 
\end{equation} 

\begin{equation} 
C \approx -0.023 - 0.005 \Omega^{2} 
\end{equation} 

where $\Omega$ is the ratio of the rotational period
of the companion star to the orbital period of the system.
Note that, whilst $R_{\rm L}/a^{\prime}$ is expected
to be constant for a given system, $R_{\rm L}$, $a^{\prime}$
and $\beta$ will vary around the orbit for a system which has 
an appreciable eccentricity.

Only seven eclipsing X-ray binary pulsars are known (namely Her X-1, Cen X-3,
Vela X-1, SMC X-1, LMC X-4, QV Nor and OAO1657--415). Orbital parameters for 
the first six of these are still relatively poorly known, whilst the
counterpart to OAO1657--415 has only recently been identified (Chakrabarty et 
al 2002) and no optical radial velocity curve has been measured. In addition,
the mass of the neutron star in a seventh eclipsing X-ray binary (4U1700--37) 
has recently been determined by Clark et al (2002). This system does not
contain an X-ray pulsar though, and the mass determination is based on a 
Monte-Carlo modelling method which relies on the spectral type of the 
companion and is thus highly uncertain. Vela X-1 is the only one of these 
systems to have an eccentric orbit, and apart from OAO1657--415 is the one 
with the longest orbital period.  For these reasons, and others discussed 
below, an accurate determination of the mass of the neutron star in Vela X-1 
has always been difficult to obtain.

\section{The history of Vela X-1}

Vela X-1 was first detected by a rocket-borne experiment (Chodil et al 1967)
and subsequent observations (see, for example, Giacconi et al 1972), suggested
that the source was highly variable. Using data from the {\em OSO-7} satellite,
Ulmer et al (1972) demonstrated evidence for intensity variations which were
interpreted as eclipses with a period of around 9 days. An optical counterpart,
GP~Vel / HD~77581 (a B0.5 giant with $m_v = 6.8$)  was identified by Brucato \&
Kristian (1972) and Hiltner et al (1972), based on its ultra-violet excess and
radial velocity variations.

An X-ray pulse period of 283~s was subsequently discovered using the {\em
SAS-3} satellite (Rappaport \& McClintock 1975, McClintock et al 1976). Timing
observations of these pulses by Rappaport, Joss \& McClintock (1976) allowed
the radial velocity semi-amplitude of the X-ray component to be measured as
$K_{\rm x} = 273 \pm 9$~km~s$^{-1}$, and the eccentricity of the system, $e
\sim 0.1$. Later work, using {\em Hakucho} and {\em Tenma} data (Deeter et al
1987), gave a value of $a_{\rm x} \sin i = 113.0\pm0.4$ light seconds for the
projected semi-major axis. Using values for the orbital period $P$ and the
eccentricity of the orbit $e$, the corresponding $K_{\rm x}$ value may be
calculated according to

\begin{equation}
K_{\rm x} = \frac{2\pi}{P} \frac{a_{\rm x} \sin i}{(1-e^2)^{1/2}}
\end{equation}

as $K_{\rm x} = 276 \pm 1$~km~s$^{-1}$.  The most recent, and accurate, values
determined from X-ray pulse timing analysis are those obtained with {\em BATSE}
on board {\em CGRO} reported by Bildsten et al (1997). They obtain a value
$a_{\rm x} \sin i = 113.89 \pm 0.13$ light seconds which corresponds to $K_{\rm
x} = 278.1 \pm 0.3$~km~s$^{-1}$ when combined with their accurate values for
the orbital period, $P = 8.964368 \pm 0.000040$~d, and eccentricity, $e =
0.0898 \pm 0.0012$.

The X-ray eclipse duration appears to be quite variable, and somewhat energy
dependent. For example, Forman et al (1973) obtained a value for the 
half angle of the eclipse of $\theta_{e} = 38^{\circ}\pm1^{\circ}$ using the 
{\em Uhuru} satellite and Charles et al (1976) obtained a value of 
$\theta_{e} = 39^{\circ}.8\pm0^{\circ}.4$ using the {\em Copernicus} satellite,
whereas Watson \& Griffiths (1977) quote a value of $\theta_{e} = 
33^{\circ}.8\pm1^{\circ}.3$ obtained using {\em Ariel V} data. However the 
earlier two experiments were at much softer energies than Watson \& Griffiths 
observed, and softer X-rays are much more likely to be absorbed by 
circumstellar material, thus extending the observed eclipse time.

\begin{table}
\caption{Measured amplitudes of the radial velocity curve of GP~Vel}
\begin{tabular}{ll} \hline
$K_{\rm o}$ / km~s$^{-1}$ & Reference \\ \hline
37--45		& Wallerstein (1974) \\
$26\pm0.7$	& Zuiderwijk 1974 \\
$20\pm1$	& van Paradijs et al (1976) \\
$21.75\pm1.15$  & van Paradijs et al (1977) \\
$21.8\pm 1.2$	& Rappaport \& Joss (1983) \\
18.0--28.2	& (95\% conf. range) van Kerkwijk et al (1995) \\
$17.8\pm1.6$	& Stickland et al (1997) \\
22		& (Correction to the above) Barziv et al (2001) \\
$21.7\pm1.6$	& Barziv et al (2001) \\
$22.6\pm1.5$    & this paper \\ \hline
\end{tabular}
\end{table}

Early determinations of $K_{\rm o}$ were made by Wallerstein (1974) and
Zuiderwijk et al (1974), see Table 1. As regular X-ray pulsations had yet to
be discovered at this point, assumptions had to be made about the mass of the
optical component in order to estimate the mass of the compact object. 
Zuiderwijk et al (1974) obtained a value of $M_{\rm x}>2.5\pm0.3$~M$_{\odot}$,
and suggested that such a large mass coupled with the lack of regular
pulsations indicated that the compact object was a black hole.  The discovery
of regular X-ray pulsations provided a means of determining the masses of both
components directly, and also ruled out the possibility of the compact object
being a black hole.  

Van Paradijs et al (1976) combined their $K_{\rm o}$ value obtained from 26
coud\'{e} spectrograms (Table 1) with the $K_{\rm x}$ value of Rappaport \&
McClintock (1975). Using X-ray eclipse data, they determined that $i >
74^{\circ}$, and thus arrived at a mass of $1.6 \pm 0.3$~M$_{\odot}$ for the
neutron star. Van Paradijs et al (1977) subsequently refined their  $K_{\rm o}$
value using yet more photographic spectra (Table 1) and Rappaport \& Joss
(1983) revised $K_{\rm o}$ further (Table 1) obtaining a neutron star mass
estimate of $1.85^{+0.35}_{-0.30}$~M$_{\odot}$ by combining data from a number
of sources, including Watson \& Griffiths (1977), and Rappaport, Joss \&
Stothers (1980), and performing a Monte Carlo analysis to estimate the
uncertainties.

More recently, van Kerkwijk et al (1995) made further optical observations of
GP Vel, and discovered strong deviations from a pure Keplerian velocity curve,
which were auto-correlated within a single night, but not from one night to
another. It was suggested that the variable gravitational force exerted by the
neutron star as it travels around its eccentric orbit excites short-lived
oscillations on the surface of the optical component which affect the measured
radial velocity. From their $K_{\rm o}$ value (Table 1) van Kerkwijk et al
(1995) obtained $M_{\rm x}=1.9^{+0.7}_{-0.5}$~M$_{\odot}$.  A significantly
lower value for $K_{\rm o}$ (Table 1) was obtained from observations using the
{\em IUE} satellite by Stickland, Lloyd \& Radziun-Woodham (1997). However,
Barziv et al (2001) report that the analysis of these {\em IUE} data was
subject to an error and a correct analysis yields a value consistent with those
previously measured (Table 1) thus solving the discrepancy.

The most recent measurement of the optical radial velocity curve of GP Vel
(Barziv et al 2001) made use of 183 spectra obtained over a nine month 
campaign in order to try to average out the deviations reported by van
Kerkwijk et al (1995). Although they were quite successful in averaging
out these excursions, they were left with different, phase-locked deviations
in the radial velocity curve. Despite this they determined an accurate 
$K_{\rm o}$ value (Table 1) and set a limit on the neutron star mass of 
$M_x \sin^3 i = 1.78 \pm 0.15$~M$_{\odot}$.

\section{Observations}

To take account of the deviations from Keplerian motion
that are present in the radial velocity curve of GP Vel, we adopted a
somewhat different approach to that employed by Barziv et al (2001). 
Instead of trying to average the velocity excursions over many orbital
periods, we chose to carry out a comprehensive radial velocity study
with maximum phase coverage over several consecutive orbital cycles.
In this way we aimed to track the velocity excursions closely and thus
model and ultimately remove them.

The observations of GP~Vel reported here were made using the 74-inch telescope
at Mount Stromlo Observatory, Canberra, Australia.  In order to cover two
complete orbits, twenty-one continuous nights were obtained
between 1996 February 1 and 1996 February 21.  Only two nights were completely
lost (the 9th and 17th) although most of the first night was spent trying
different setup configurations, so this too produced little data.  Typically,
four sets of three spectra were obtained on each of the remaining nights, with
exposures mostly of 1000s. In total 180 spectra of GP~Vel were obtained. 

The echelle and CCD camera were mounted at the coud\'{e} focus. All
observations used the 31.6 lines mm$^{-1}$ echelle grating, the 158 lines
mm$^{-1}$ cross disperser grating, the 81.3~cm focal length camera and the
thinned 2k~$\times$~2k Tektronix CCD, with a slit width of 2 arcsec.  Around 70
separate echelle orders fell onto the detector spanning a wavelength range 3370
-- 5970 \AA. To shorten the readout time, $2\times2$ binning was used on the
CCD for the first seven  nights, but this was changed to $2\times1$ binning 
(i.e. binning only in the cross-dispersion direction) for the remainder of 
the run. 

The radial velocity standards HR~1829 (a G5II star) and HR~3694 (a K5III--IV
star) were also observed on most nights, as well as a B0Ia star
($\epsilon$~Ori) to provide a template spectrum for the cross correlation. We
also obtained the usual complement of flat fields and bias frames as well as
thorium-argon arcs each time the telescope was moved.

\section{Data Reduction and Analysis}

The {\sc iraf} software package was used to perform the reduction and 
analysis. For each night's spectra, an average bias frame was constructed 
and subtracted from the arc, flat and object frames. Orders were traced
automatically using the GP~Vel frames, as the object was relatively bright 
and its early-type spectrum does not contain many large absorption features 
which could complicate the tracing process. An optimal extraction algorithm 
was used to extract all the spectra in order to maximise the
signal-to-noise ratio.

A wavelength calibration solution was found for the whole echelle frame using
the arcs. A total of 1109 lines were identified from the thorium-argon arc
spectrum (about 20 per order of the echelle) and the wavelength solution was
fitted using a Chebyshev polynomial of order three in the wavelength direction
plus another of order four in the spatial direction. The RMS of the fit was
0.005\AA, which is less than 10\% of a pixel in all orders. The FWHM of the
individual arc lines near the centre of the echelle about 0.1\AA.  

\begin{figure*}[ht]
\setlength{\unitlength}{1cm}
\begin{picture}(12,21)
\put(-1,-3){\includegraphics{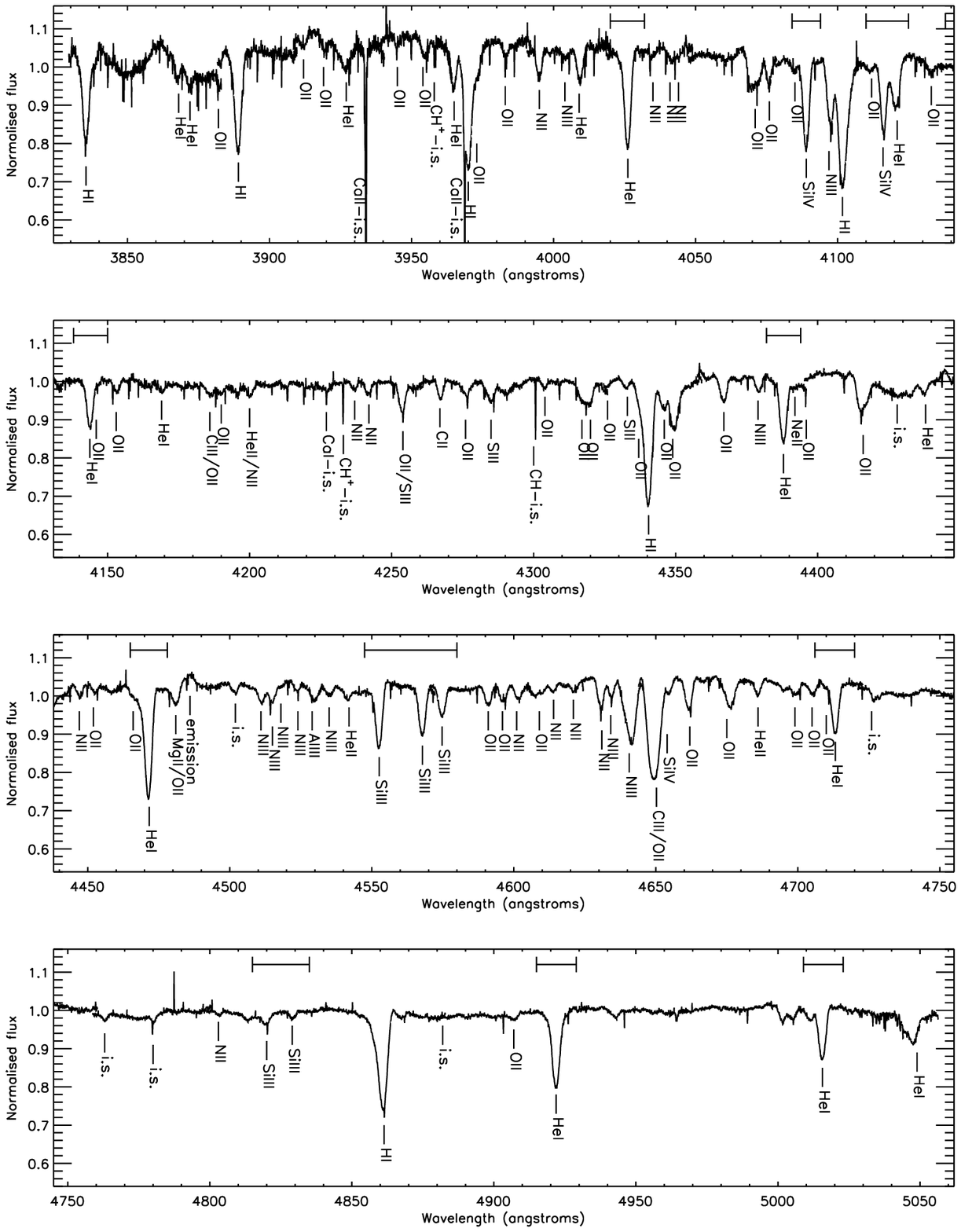}}
\end{picture}
\caption{An average spectrum of GP Vel. Horizontal bars indicate the regions
used in the cross-correlation procedure. Many of the sharp lines peaking 
downwards are weak interstellar lines; most of the upward peaking lines
are remaining sky features.}
\end{figure*}

\clearpage

At the blue end, the signal-to-noise ratio
was poor and there was also a problem with overlapping orders.  Consequently
only 53 orders were extracted from each echelle spectrum, spanning the
wavelength range 3820\AA \ to 5958\AA \ (orders 148 to 96). The spectral
resolution ranged from 0.06\AA \ pixel $^{-1}$ at the blue end to 0.1\AA \
pixel $^{-1}$ at the red end. All spectra were extracted onto a common
log-linearised scale with a resolution of 4.86~km~s$^{-1}$ per pixel.  
An average spectrum is shown in Figure 1.

The spectra of GP~Vel showed essentially the same set of absorption lines 
as were seen by van Kerkwijk et al (1995). In particular we saw the Balmer 
series from H$\beta$ to H9, He{\sc i} lines at 5015\AA, 4921\AA, 4713\AA,
4471\AA, 4387\AA, 4143\AA, 4121\AA \ and 4026\AA, and several lines due to
Si{\sc iii} (4553\AA, 4568\AA, 4575\AA, 4820\AA, 4829\AA), Si{\sc iv}
(4089\AA, 4116\AA), O{\sc ii} (4591\AA, 4596\AA) and N{\sc ii} (4601\AA).

Before proceeding with the main cross-correlation, the stability of the
detection system was checked by cross-correlating the various spectra of the
radial velocity standards HR~1829 and HR~3694 against each other. Across the
whole run, the mean shift in these spectra was only 0.46~km~s$^{-1}$, which is
negligible ($<0.1$ of a pixel). As a further check, regions of the GP~Vel
spectra containing the CaII 3934\AA \ and NaI 5890/5896\AA \ interstellar lines
were cross correlated against themselves. No trends were apparent in these data
either.

To extract the radial velocity curve of GP~Vel, template spectra of
$\epsilon$~Ori were cross correlated against the spectra of the target. Since
these stars have similar spectral types, systematic errors in this process
should be minimised. Regions of the spectrum containing the He{\sc i} lines
plus the Si{\sc iii} and Si{\sc iv} lines were used to produce the radial
velocity curve in Figure 2 (see Appendix for the individual radial 
velocity values, relative to $\epsilon$ Ori). The radial velocities 
themselves were extracted using standard {\sc iraf} routines by
fitting a Gaussian to the cross-correlation function in each case.  The 
Gaussian fits were performed using the full shape of the Gaussian curve
down to one-quarter of their height. Although
radial velocities were also obtained from the Balmer lines, these lines were in
general very broad and so it was difficult to obtain accurate velocities.
Furthermore, there was some evidence that the radial velocity amplitude
increased with the order of the Balmer line. We comment on this further below,
but as a result did not use Balmer lines in the rest of our analysis.

\begin{figure}
\setlength{\unitlength}{1cm}
\begin{picture}(6,7)
\put(-0.5,7){\includegraphics{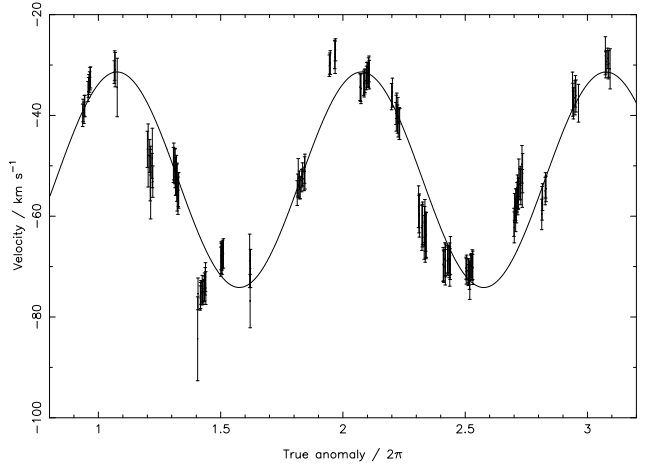}}
\end{picture}
\caption{Radial velocity curve for He{\sc i}, Si{\sc iii} and Si{\sc iv} 
lines showing the fitted curve (the first fit). Each cluster of radial 
velocity values represents a single night's data. The fit has a reduced
chi-squared of 3.84.}
\end{figure}

\section{Fitting the radial velocity curve}

Because Vela X-1 is known to have a significant eccentricity, simply fitting
a sinusoidal radial velocity curve to the cross-correlation results as a
function of time would not be satisfactory, as an eccentric orbit will show
deviations from a pure sinusoid. Instead it may be shown that the 
observed radial velocity is given by:

\begin{equation}
v = \gamma + \frac{4\pi a_1 \sin i}{P(1-e^{2})^{\frac{1}{2}}}
 \frac{e\cos \omega + \cos(\nu + \omega)}{2}
\end{equation}

where $a_1$ is the semi-major axis of the orbit of GP~Vel, $e$ is the orbital
eccentricity, $P$ is the orbital period, $\nu$ is the true anomaly (i.e. the
angle between the major axis and a line from the star to the focus of the
ellipse), $\omega$ is the periastron angle (i.e. the angle between a line from 
the centre of the orbit to periastron and a line from the centre of the orbit 
to the ascending node), $i$ is the angle between the normal to the plane
of the orbit and the line-of-sight and $\gamma$ is the radial velocity of the
centre of mass of the binary system. The true anomaly is related to the
eccentric anomaly, $E$, by:

\begin{equation}
\tan\left(\frac{\nu}{2}\right)=\left(\frac{1+e}{1-e}\right)^{\frac{1}{2}}
\tan \left(\frac{E}{2} \right)
\end{equation}

where $E$ is the angle between the major axis of the ellipse and the line 
joining the position of the object and the centre of the ellipse. In turn, 
$E$ can be related to $M$, the mean anomaly, by Kepler's Equation:

\begin{equation} 
E-e \sin E = M = \frac{2 \pi}{P}(t-T_0)
\end{equation}

where the right hand side of this equation is simply the orbital phase
with $T_0$ the time of periastron passage.
This equation can be solved numerically, or using Bessel functions.

The value of $T_0$ was obtained from the value of $T_{\pi/2}$, the epoch
of 90$^{\circ}$ mean orbital longitude, derived by Bildsten et al (1997)
from {\em BATSE} pulse timing data. They give $T_{\pi/2}$ = MJD 48895.2186
$\pm$ 0.0012, from which

\begin{equation}
T_0 = T_{\pi /2} + \frac{P \left( \omega - \frac{\pi}{2} \right)}{2\pi}
\end{equation}

where $\omega = 152.59^{\circ} \pm 0.92^{\circ}$. Hence we calculate the 
time of periastron passage as $T_0$ = MJD 48896.777 $\pm$ 0.009.

\subsection{The first fit}

Times, $t$, were assigned to each radial velocity measurement and
Equation 12 was solved using a numerical grid to calculate the corresponding
eccentric anomaly. The true anomaly was then found using Equation 11, and
Figure 2 shows a plot of radial velocity against true anomaly, with the best
fit curve according to Equation 10.  The reduced chi-squared of the fit
is $\chi^2_{\rm r} = 3.84$. Scaling the error bars by a factor of 1.96 to
reduce $\chi^2_{\rm r}$ to unity, gives the amplitude of the fitted curve
as $K_{\rm o} = 21.4 \pm 0.5$~km~s$^{-1}$. However, we note
that, since the use of chi-squared assumes that the errors on all the points 
are uncorrelated, the uncertainty here is likely to be a gross under-estimate.
This will hereafter be referred to as the `first fit'.

Figure 3 shows the residuals to the radial velocity curve in the first fit,
plotted against time. It is clear that there are trends apparent in these
residuals from night to night and a Fourier analysis shows that the dominant
signals are at periods of $9 \pm 1$~d and $2.18 \pm 0.04$~d (Figure 4). 
The 9~d signal present in the power spectrum reflects the fact that fixing 
the zero phase of the radial velocity curve, as implied by the X-ray data of 
Bildsten et al (1997) does not provide the best fit to the data. We suggest 
this effect may be responsible for the `phase-locked' deviations revealed 
by Barziv et al (2001) too. The 2.18~d modulation appears to be relatively 
stable throughout our two orbits of observations, with an amplitude of around
5~km~s$^{-1}$, as shown in Figure 3.

\begin{figure}
\setlength{\unitlength}{1cm}
\begin{picture}(6,7)
\put(0,7){\includegraphics{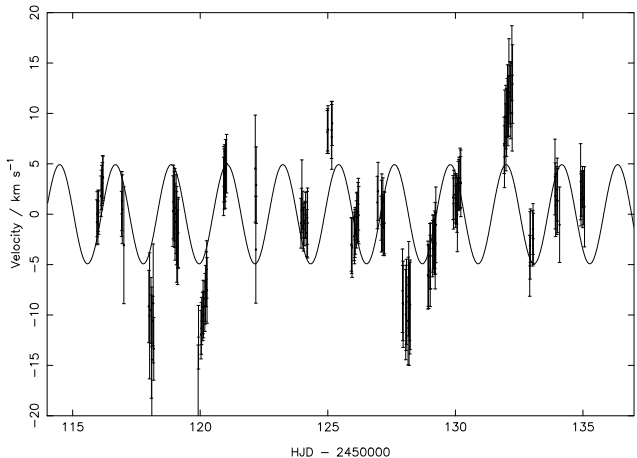}}
\end{picture}
\caption{Residuals to the radial velocity curve fit in Figure 2,
plotted against time. (The first fit.) Overlaid is a best-fit sinusoid with
a period of 2.18~d.}
\end{figure}

\begin{figure}
\setlength{\unitlength}{1cm}
\begin{picture}(6,7)
\put(0,7){\includegraphics{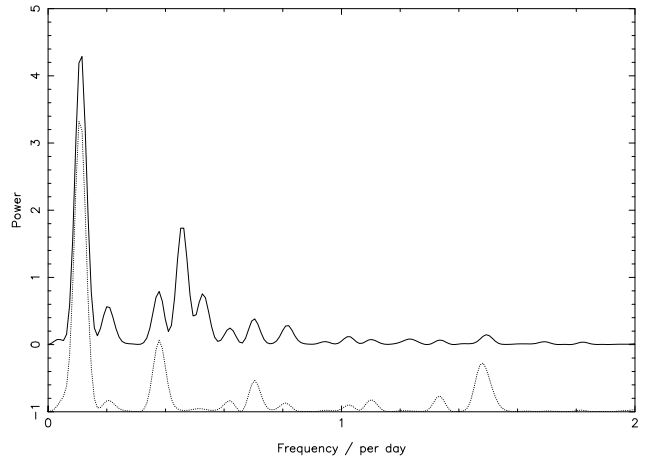}}
\end{picture}
\caption{Power spectra of the residuals to the radial velocity curve fits.
The solid curve is the power spectrum of the data in Figure 3 (the residuals
after the first fit) and the dotted curve (offset in the $-$ve direction
by one unit for clarity) is the power spectrum of the data
in Figure 6 (the residuals after the second fit). The highest peaks in 
the power spectrum of the residuals after the first fit correspond to
frequencies of $0.111~{\rm d}^{-1}$ and $0.459~{\rm d}^{-1}$ or periods 
of 9.0~d and 2.18~d. In each case, window function peaks due to the 
sampling of the lightcurve have been removed using a 1-dimensional 
clean (program courtesy H. Lehto).}
\end{figure}

\subsection{The second fit}

In order to take account of the correlated residuals from the first fit,
another fit to the radial velocity data was performed this time with four free
parameters: the $\gamma$ velocity and the $K_{\rm o}$ value as before, plus the
amplitude and phase of a sinusoid with a period of 2.18~d.  The $K_{\rm o}$
part of the fit is sinusoidal as a function of true anomaly, whilst the 2.18~d
part of the fit is sinusoidal purely as a function of time. The radial velocity
fit according to this prescription, referred to as the second fit, is shown in
Figure 5 and has a reduced chi-squared of 2.40. The improvement over the first
fit is therefore clearly apparent. Scaling the error bars by a factor of 1.55 
to reduce $\chi^2_{\rm r}$ to unity, gives the amplitude of the fitted curve
as $K_{\rm o} = 22.4 \pm 0.5$~km~s$^{-1}$ and the amplitude of the 2.18~d
modulation as $5.4 \pm 0.5$~km~s$^{-1}$. As with the first fit, we note
that the uncertainties here are also likely to be under-estimates. The 
remaining residuals on the second fit are shown in Figure 6, and their power 
spectrum is indicated by the dotted line in Figure 4. It can be seen that some 
of the correlated structure in the residuals has been removed by this
procedure, although the peak corresponding to a period of $\sim 9$~d is 
still present indicating once again that the zero phase according to the X-ray 
observations does not provide the best description.

\begin{figure}
\setlength{\unitlength}{1cm}
\begin{picture}(6,7)
\put(-0.5,7){\includegraphics{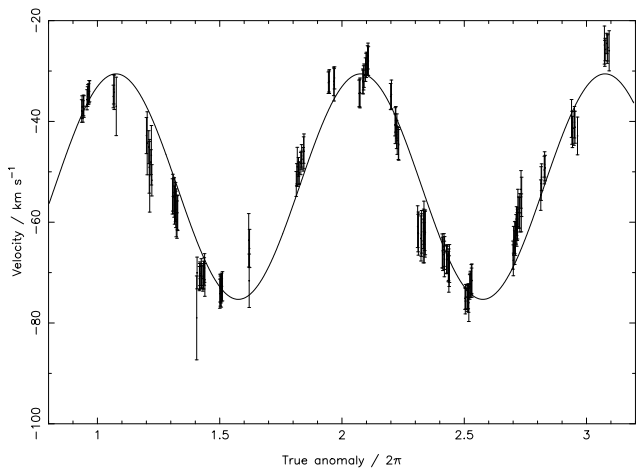}}
\end{picture}
\caption{Radial velocity curve for He{\sc i}, Si{\sc iii} and Si{\sc iv} 
lines showing the fitted curve (the second fit) after removal of the 2.18 
hour signal.  Each cluster of radial 
velocity values represents a single night's data. The fit has a reduced
chi-squared of 2.40.}
\end{figure}

\begin{figure}
\setlength{\unitlength}{1cm}
\begin{picture}(6,7)
\put(0,7){\includegraphics{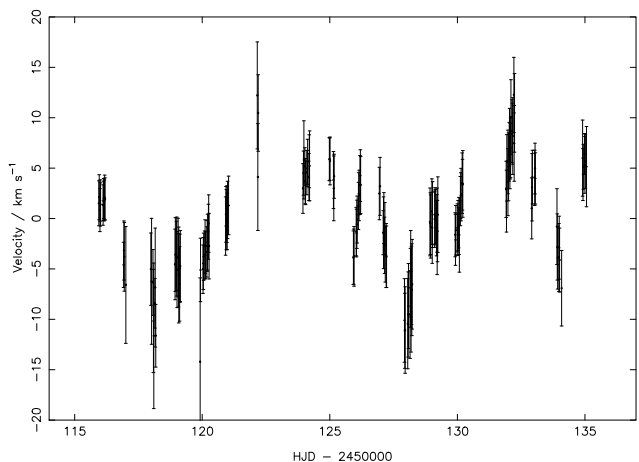}}
\end{picture}
\caption{Residuals to the radial velocity curve fit in Figure 5,
plotted against time. (The second fit.)}
\end{figure}

\subsection{Fits to phase binned means}

To test whether the deviations from the radial velocity curve 
displayed by each group of data points are correlated, we subtracted the
effect of the 2.18~d period according to the parameters found for it 
in the second fit, and rebinned the resulting radial velocities into nine 
phase bins (each $\sim 1$ day). The radial velocity fit to these phase-binned
means is shown in Figure 7 and has a reduced chi-squared of 28.3. The errors
on the mean data points are calculated as the standard error on the mean 
from the individual data points that are averaged in each case. The fact that 
the reduced chi-squared is significantly greater than unity suggests that 
the deviations are indeed correlated. Scaling the error bars here by a factor 
of 5.3 to reduce $\chi^2_{\rm r}$ to unity, gives the amplitude of the fitted 
curve as $K_{\rm o} = 22.6 \pm 1.5$~km~s$^{-1}$. 

Figure 7 clearly shows that a better fit would be obtained if the zero phase of
the radial velocity is allowed as a free parameter.  This reflects the fact
that a $\sim 9$~d period is present in the power spectra of the residuals after
both the first and second fits (Figure 4). Allowing the zero phase to vary
results in the second fit to the phase binned data shown by the dashed 
line in Figure 7. This time the reduced chi-squared is 6.0. 
Scaling the error bars here by a factor of 2.44 to reduce $\chi^2_{\rm r}$ to
unity, gives the amplitude of the fitted curve as $21.2 \pm
0.7$~km~s$^{-1}$. The shift of the minimum radial velocity with respect to 
that in the first fit to the pase binned means is a true anomaly of $\Delta 
\nu = 0.033 \pm 0.007$, corresponding to about 7
hours. We note that this shift in zero phase is significantly greater than the
uncertainty implied by extrapolating the X-ray ephemeris of Bildsten et al
(1997) to the epoch of our observations which is only $\sim 0.0006$ of a cycle.
We conclude that the orbital phase locked deviations remaining in
the radial velocities after the first fit to means can therefore 
mostly be accounted for by a simple phase shift in the radial velocity curve.

\begin{figure}
\setlength{\unitlength}{1cm}
\begin{picture}(6,7)
\put(-0.5,7){\includegraphics{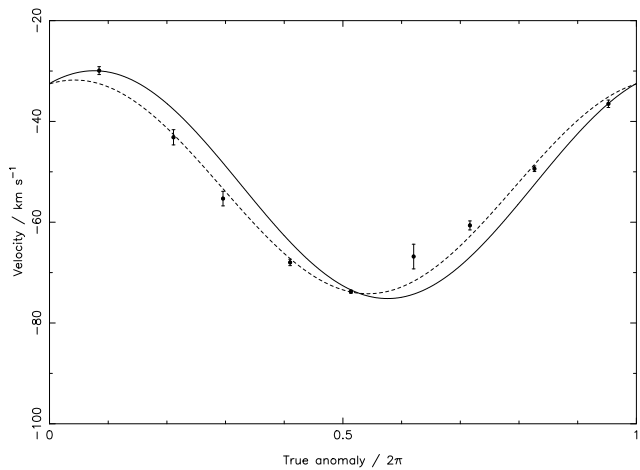}}
\end{picture}
\caption{Radial velocity curve for He{\sc i}, Si{\sc iii} and Si{\sc iv} 
lines binned into nine phase bins. Each radial velocity value represents 
two or more night's data. The fit to the phase binned means has a reduced 
chi-squared of 28.3 and is shown by the solid curve. A second fit to the 
phase binned means, allowing the zero phase of the radial velocity as a free
parameter, is shown by the dashed line and has a reduced chi-squared of 6.0.}
\end{figure}

\section{Calculation of system parameters}

The masses of the two stellar components were calculated according to the method
described in Section 1. In order to propagate the uncertainties, a Monte Carlo
approach was adopted, calculating $10^6$ solutions with input values drawn from
a Gaussian distribution within the respective 1$\sigma$ uncertainties of each
parameter. The final values were then calculated as the mean result of these
$10^6$ solutions, with the 1$\sigma$ uncertainties given by the rms deviation
in each case.

The procedure was carried out using $K_{\rm o} = 22.6 \pm
1.5$~km s$^{-1}$ from the first fit to means and the results of these
calculations are shown in Table 2. We used the value of $K_{\rm x} = 278.1
\pm 0.3$~km s$^{-1}$ (Bildsten et al 1997), to calculate $q$ using Equation 1
and a value of $33^{\circ}.8\pm1.3^{\circ}$ was adopted for the eclipse
half-angle, $\theta_{e}$ from Watson \& Griffiths (1977). We note that this
value was also used by Wilson \& Terrel (1998) for their unified analysis of
Vela X-1 and appears to be the most reliable. A value of $0.67\pm0.04$ was
adopted for the co-rotation factor, $\Omega$, obtained from comparison of
GP~Vel spectra with model line profiles by Zuiderwijk (1995). We note 
that Barziv et al (2001) also used the $v \sin i$ value derived by Zuiderwijk
(1985) and obtained a similar value, namely $\Omega (=f_{\rm co}) = 
0.69 \pm 0.08$.  

Values for  $\beta$ and $i$ cannot be obtained from independent observations. 
However, since the supergiant is unlikely to be overfilling its Roche lobe, an
upper limit to the Roche lobe filling factor is $\beta \leq 1.0$; and an upper
limit to the inclination is clearly $i \leq 90^{\circ}$. A lower limit to
$\beta$ is found by setting $i=90^{\circ}$, and a lower limit to $i$ is found by
setting $\beta = 1$. These limits are indicated in Table 2. We note though that
$\beta$ is unlikely to have a fixed value, as the size and shape of the Roche
lobes of both components will vary as the separation between the stars varies
as they orbit each other eccentrically. Indeed, it has been suggested that
GP~Vel is likely to fill its Roche lobe at periastron (Zuiderwijk 1995).
However the mass ratio implies that Roche lobe overflow would be dynamically
unstable and lead to a common-envelope phase, so the limit of $\beta = 1$ is
probably implausible. We also note that Barziv et al (2001) took a somewhat
different approach and assumed a filling factor at periastron in the range
$\beta = 0.9 - 1.0$, thus leading to a minimum value for the inclination of
$73^{\circ}$, consistent with our analysis.

Given these two limits on each of $\beta$ and $i$, Equations 2 and 3 then allow
us to calculate the masses of the two components for combinations of $i$ and
$\beta$ between the two extremes. The limiting values for the masses of the
two stars are shown in Table 2. Values between these extremes, as a function of
inclination angle, are shown in Figure 8.  With these two limiting situations,
we also calculate the semi major axis $a$ of the orbit from Kepler's law,
and this is then used to calculate the separation of the stars at periastron
according to $a^{\prime} = a (1-e)$. The Roche lobe radius at periastron
$R_{\rm L}$ is then determined from the value of $R_{\rm L}/a^{\prime}$ at each
extreme, and the radius of the companion star $R_{\rm o}$ is then calculated as
$\beta \times R_{\rm L}$ in each case. These values too are shown in Table 2.
For each parameter, the actual value may be anywhere within the range
encompassed by the two limits.

\begin{table}
\caption{\bf System parameters for Vela X-1 / GP~Vel. The value for $K_{\rm o}$ is
that resulting from fitting the phase bin means without a phase shift.}
\begin{tabular}{llll} \hline
Parameter & \multicolumn{2}{c}{Value} & Ref. \\ \hline
{\it Observed}     		&	&	& \\
$a_{\rm x} \sin i$ / lt sec	& \multicolumn{2}{c}{$113.98 \pm 0.13$}		& [1]\\
$P$ / d				& \multicolumn{2}{c}{$8.964368 \pm 0.000040$}	& [1]\\
$T_{\pi/2}$ / MJD		& \multicolumn{2}{c}{$48895.2186 \pm 0.0012$}	& [1]\\	
$e$				& \multicolumn{2}{c}{$0.0898 \pm 0.0012$}	& [1]\\	
$\omega$ / deg			& \multicolumn{2}{c}{$152.59 \pm 0.92$}		& [1]\\
$\theta_{\rm e}$ / deg		& \multicolumn{2}{c}{$33.8 \pm 1.3$}		& [2]\\	
$\Omega$			& \multicolumn{2}{c}{$0.67 \pm 0.04$}		& [3]\\
$K_{\rm o}$ / km s$^{-1}$	& \multicolumn{2}{c}{$22.6 \pm 1.5$} 		& [4]\\
{\it Inferred} & & & \\
$K_{\rm x}$ / km s$^{-1}$	& \multicolumn{2}{c}{$278.1 \pm 0.3$}  		& \\
$T_0$ / MJD			& \multicolumn{2}{c}{$48896.777 \pm 0.009$} 	& \\
$q$				& \multicolumn{2}{c}{$0.081 \pm 0.005$}  \\
$\beta$				& $1.000$ 	& $0.89 \pm 0.03$	 \\
$i$ / deg			& $70.1 \pm 2.6$& $90.0$                 \\
$M_{\rm x}$ / M$_{\odot}$ 	& $2.27 \pm 0.17$ & $1.88 \pm 0.13$ 	 \\
$M_{\rm o}$ / M$_{\odot}$ 	& $27.9 \pm 1.3$& $23.1 \pm 0.2$ 	 \\
$a^{\prime}$ at periastron / R$_{\odot}$ & $51.4 \pm 0.8$  & $48.3 \pm 0.3$  \\
$R_{\rm L}$ at periastron / R$_{\odot}$	& $32.1 \pm 0.6$  & $30.2 \pm 0.2$   \\
$R_{\rm o}$ / R$_{\odot}$	& $32.1 \pm 0.6$  & $26.8 \pm 0.9$           \\ \hline
\end{tabular}\\
$[1]$ Bildsten et al 1997\\
$[2]$ Watson \& Griffiths 1977\\
$[3]$ Zuiderwijk 1995 \\
$[4]$ this paper 
\end{table}

\begin{figure}
\setlength{\unitlength}{1cm}
\begin{picture}(5,13)
\put(0,0){\includegraphics{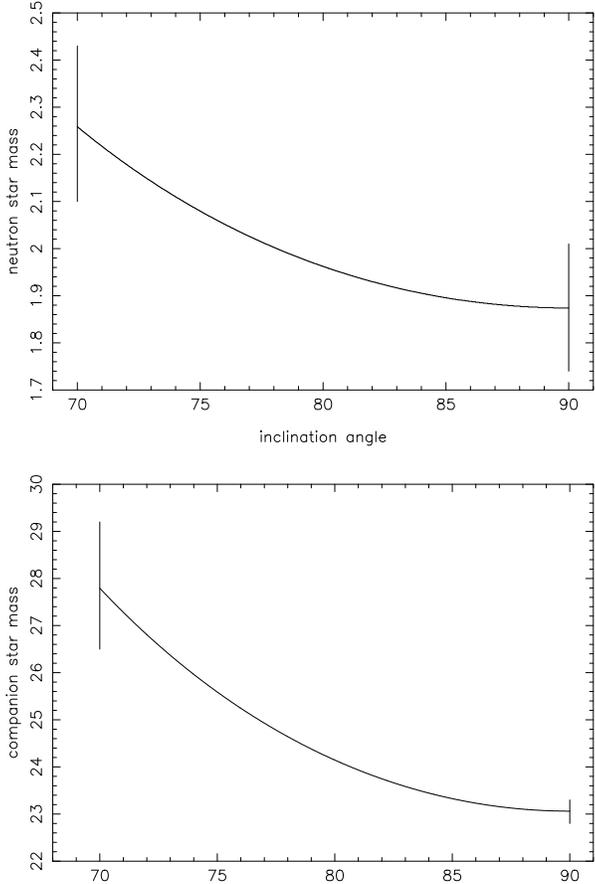}}
\end{picture}
\caption{The masses of the neutron star (upper panel) and companion star 
(lower panel) in solar units, as a function of inclination angle. The lower 
limit to the inclination angle is set by the condition 
that the companion star cannot over-fill its Roche lobe. Error bars on the 
masses at the extreme limits are determined as explained in the text.}
\end{figure}

\section{Tidally induced non-radial oscillations}

The strongest peaks in the power spectrum of the residuals after the
first fit, at 0.111~d$^{-1}$ and 0.457~d$^{-1}$, are suggestively close to the
orbital frequency and $4\times$ the orbital frequency ($f_{\rm orb} =
0.11155\,{\rm d}^{-1}$) respectively. This leads us to speculate that they may
both represent tidally excited non-radial oscillation modes (see also
Willems \& Aerts 2002, Handler et al. 2002).  

From a theoretical point of view, the tidal action exerted by one
binary component on the other is governed by the tide-generating
potential which can be expanded in a Fourier series in terms of multiples 
of the mean motion $n=2\,\pi\,f_{\rm orb}$ (e.g. Polfliet \& Smeyers 1990). The 
oscillations that are most likely to be excited are those associated with 
the lower-order harmonics of the
Fourier series. In the particular case of GP~Vel, the dominant
contributions to the Fourier series of the tide-generating potential
are associated with the first ten harmonics. This supports the
possibility that the oscillation associated with the frequencies 
$\approx  f_{\rm orb}$  and $\approx 4\, f_{\rm orb}$ may be induced 
by the tidal action of the neutron star.

An in-depth analysis of the possibility that a tidally excited
oscillation mode exists in GP~Vel requires an accurate knowledge of
the stellar and orbital parameters. In particular, the internal
structure of the B-type companion to the neutron star must be known to
derive the spectrum of the eigenfrequencies present in the star. The
radii quoted in Table~2 indicate that the B-type star might be in 
the Hertzsprung-gap. An additional complication arises from the
unknown evolutionary history of the binary as the B-type star may be
`polluted' by mass-transfer episodes prior to the formation of the
neutron star. In view of these complications, a detailed study of the
oscillation spectrum of GP~Vel and its possible use in an
asteroseismological study is beyond the scope of our present
investigation.

\section{Discussion}

Our final value for $K_{\rm o}$ of $22.6 \pm 1.5$~km~s$^{-1}$ is slightly
higher than, but comparable to, those quoted by other recent authors, noted in
Section 2. Our result for the neutron star mass is compatible with van Kerkwijk
et al's (1995) mass value of $M_{\rm x}=1.9^{+0.7}_{-0.5}$M$_{\odot}$, and that
of Barziv et al (2001), namely  $M_{\rm x}=1.86 \pm 0.32$M$_{\odot}$.  We
therefore support the view that the neutron star in Vela X-1 is more massive
than the canonical value of 1.4M$_{\odot}$.  However, there are a number of
potential sources of systematic error, as noted in our earlier paper concerning
the mass of the neutron star in Cen X-3 (Ash et al 1999), some of which will be
examined below.

\subsection{Stellar wind}

As GP~Vel is an early-type star, it has a significant stellar wind, and a
strong and variable stellar wind can have disastrous consequences when making
radial velocity measurements, especially over an extended period of time. 
Barziv et al (2001) made a detailed study of the effects of winds
on their observations and showed that deviations in the Balmer lines in 
particular indicated the presence of a photo-ionization wake. We note 
that the supergiant GP~Vel is of later type (B0) than that
of Cen X-3 (O6-7), and thus has a weaker wind.  Additionally, as our
observations were made over a relatively short time period, the potential
effect of long-term wind variation will be less than if the observations were
spread over a number of years, as in the case of our Cen X-3 campaign, or 
over a year, as in the case of Barziv et al's Vela X-1 campaign. In other
words, although the effect of the supergiant's stellar wind cannot be
discounted, its effect should be less than its effect on our Cen X-3 results
and less than on the results of Barziv et al (2001).

The apparent increase in amplitude of the radial velocity curve seen with
increasing order in the Balmer series lines mentioned earlier, may
also be an indication of a stellar wind. Conceivably, the different order
lines have different contributions from the wind material as it moves further 
out, and this could produce the correlation seen. Since we have not used
the Balmer lines in our analysis however, this effect does not influence
our final result.

\subsection{X-ray heating}

The effect of X-ray heating of the surface of GP~Vel may be compared with
the corresponding effect in Cen X-3. The ratio of intrinsic to irradiated
fluxes may be estimated for each system as $(L_{\rm o} / 4\pi R_{\rm o}^2) /
(L_{\rm x} / 4\pi (a - R_{\rm o})^2)$. The {\em RXTE} ASM lightcurves show that
both sources have similar X-ray fluxes, but Cen X-3 is at over three times the
distance of Vela X-1 (Sadakane et al 1985; Humphreys \& Whelan 1975) so its
X-ray luminosity is at least a factor of ten higher.  The luminosities of the
supergiant stars in both systems are similar, but the radius of the star in Cen
X-3 and the separation of the supergiant and neutron star are both around three
times smaller in Cen X-3 (Ash et al 1999) than in Vela X-1. Hence the effect of
X-ray heating in Vela X-1 is of order ten times less than in Cen X-3. We were
able to rule out significant X-ray heating in the case of Cen X-3 (Ash et al
1999), and the effect should therefore be even less in this case.

\subsection{Deviations from Keplerian radial velocity curve}

Clearly, another source of uncertainty in the case of GP~Vel arises from the
observed deviations from a pure Keplerian radial velocity curve. van Kerkwijk
et al (1995) suggested that these deviations do not last longer than a single
night, although our results (Figure 3) suggest that they may last for up
to two system orbits. They suggested that the deviations are the result of
short-lived, high-order pulsations of the photosphere and we confirm that
the frequencies appear to be harmonically related to the orbital frequency.
It is possible therefore that they do represent tidally induced oscillations.
As well as introducing another source of error into our attempts to measure the
radial velocity of the centre of mass of the supergiant, the ``wobbles'' cause
another problem.  If the shape of the star is constantly changing, this will
affect the extent to which the star fills its Roche lobe. Additionally, the
size and shape of the Roche lobe itself will vary with phase, as the two
components orbit each other in eccentric orbits. We believe that we
have partially accounted for this effect by removing the 2.18~d modulation 
shown in Figure 3.

However, as Figure 7 demonstrates, there is clearly a remaining effect
which may be characterised by an additional radial velocity variation at the
orbital period. This manifests itself as a phase shift in the radial velocity
curve with respect to the X-ray orbital ephemeris. The measured amplitude of
the radial velocity curve is therefore a combination of the orbital motion of
GP~Vel {\em and} an additional component that may be due to a non-radial
pulsation induced by the eccentric orbit of the neutron star. There may even be
another component of this non-radial pulsation at the orbital period that is
{\em in phase} with the orbital motion and so undetectable. Our $K_{\rm o}$
value, and presumably all previous measurements too, are therefore subject to
an unknown systematic error, and the stellar masses calculated from them are
also uncertain because of this.

Another possibility for the origin of the extra velocity component at the
orbital period is large-scale motion required to keep the tidal bulge pointing
in the direction of the neutron star given the asynchronous stellar rotation
($\Omega = 0.67$). However, such a tidal bulge would extend in both directions
-- towards and away from the neutron star -- and so the effect may cancel out.

As a further result, our identification of these departures from the radial
velocity curve in GP~Vel suggest that it may be the third recent candidate to
exhibit tidally induced non-radial oscillations, after HD\,177863 
(De Cat 2001, Willems \& Aerts 2002), and HD\,209295 (Handler et al. 2002).

\subsection{Neutron star mass}

The minimum neutron star mass consistent with our results is 
around 1.74~M$_{\odot}$. A recent review of neutron star equations of state by
Lattimer \& Prakash (2000) shows that such a mass would exclude the 
schematic potential models of Prakash, Ainsworth and Latimer, and the 
various models based on a field theoretical approach by Glendenning \&
Moszkowski, by Glendenning and Schaffner-Bielich, and by Prakash, Cooke
\& Lattimer. Hence the `softer' equations of state that incorporate 
components such as hyperons, kaons and quark matter are ruled out.

In order to be consistent with a neutron star mass of $\sim 1.4$~M$_{\odot}$, 
the radial velocity amplitude of GP~Vel would have to be $K_{\rm o} \sim
18$~km~s$^{-1}$, implying that the additional velocity component 
due to the non-radial pulsation has an amplitude of around 
$\sim 4.5$~km~s$^{-1}$.

We conclude by noting that there are two large sources of uncertainty in
deriving the masses of the stellar components. The first is that of the value
$\beta$ (the ratio of the radius of the companion star to its Roche lobe
radius).  This in turn leads to a large uncertainty in $\sin i$ (and hence the
masses) and is unlikely to be reduced even with optical spectroscopic
observations of greater signal-to-noise. The second source of uncertainty
is the additional radial velocity component at the orbital period that we have
discovered and represented by a phase shift. Removing the effect of such a
velocity component is likely to require detailed modelling of tidally excited
oscillation modes in GP~Vel which in turn requires an accurate knowledge of the
stellar parameters. Consequently, the mass measurement presented here may
represent the best that is achievable.

\begin{acknowledgements}
We thank the Mount Stromlo Observatory staff for maintaining the 
74-inch telescope. The data analysis reported here was carried out using 
facilities provided by PPARC, Starlink and the OU research committee.  HQ was 
employed under PPARC grant number GR/L64621 during the course
of this work. The authors would also like to thank Deepto Chakrabarty for his 
help, and the referee, Marten van Kerkwijk, for his many detailed and 
constructive suggestions which have helped to significantly improve the paper.
\end{acknowledgements}

\section*{Appendix}

\begin{table*}
\caption{The radial velocities of GP~Vel relative to that of $\epsilon$ Ori}
\begin{tabular}{ll|ll|ll}
HJD -- 2450000 & RV / km~s$^{-1}$ & HJD -- 2450000 & RV / km~s$^{-1}$ & HJD -- 2450000 & RV / km~s$^{-1}$\\ \hline
115.9492 & $-39.978 \pm 2.229$ & 123.9840 & $-52.446 \pm 3.876$ & 129.1373 & $-68.519 \pm 2.676$ \\
115.9620 & $-39.016 \pm 2.237$ & 124.0222 & $-52.899 \pm 1.868$ & 129.1820 & $-69.463 \pm 2.743$ \\
115.9900 & $-39.473 \pm 2.130$ & 124.0368 & $-54.319 \pm 2.242$ & 129.1951 & $-68.887 \pm 2.855$ \\
116.0028 & $-38.093 \pm 2.047$ & 124.0507 & $-54.361 \pm 2.167$ & 129.2080 & $-69.592 \pm 4.292$ \\
116.0153 & $-38.126 \pm 2.148$ & 124.1014 & $-52.954 \pm 2.206$ & 129.2211 & $-69.514 \pm 3.050$ \\
116.1147 & $-35.260 \pm 2.002$ & 124.1153 & $-52.398 \pm 1.780$ & 129.2344 & $-67.797 \pm 3.757$ \\
116.1273 & $-34.446 \pm 1.959$ & 124.1292 & $-51.638 \pm 2.230$ & 129.9134 & $-70.298 \pm 2.184$ \\
116.1400 & $-33.855 \pm 1.990$ & 124.1798 & $-52.602 \pm 2.338$ & 129.9264 & $-70.958 \pm 2.530$ \\
116.1643 & $-33.359 \pm 2.026$ & 124.1944 & $-50.835 \pm 2.627$ & 129.9393 & $-70.168 \pm 2.452$ \\
116.1768 & $-32.720 \pm 2.218$ & 124.2083 & $-51.151 \pm 3.466$ & 129.9995 & $-70.815 \pm 2.380$ \\
116.1905 & $-32.454 \pm 2.090$ & 124.9819 & $-30.123 \pm 2.138$ & 130.0124 & $-71.690 \pm 2.064$ \\
116.9187 & $-31.398 \pm 2.249$ & 124.9944 & $-29.702 \pm 2.094$ & 130.0251 & $-70.973 \pm 2.494$ \\
116.9314 & $-30.109 \pm 2.933$ & 125.0078 & $-29.527 \pm 2.370$ & 130.0515 & $-71.616 \pm 2.018$ \\
116.9468 & $-30.908 \pm 3.410$ & 125.1431 & $-27.910 \pm 2.830$ & 130.0650 & $-71.685 \pm 2.830$ \\
117.0059 & $-34.455 \pm 5.798$ & 125.1561 & $-28.457 \pm 3.213$ & 130.0780 & $-72.801 \pm 3.691$ \\
117.9872 & $-46.720 \pm 3.585$ & 125.1691 & $-26.982 \pm 2.189$ & 130.1179 & $-70.146 \pm 2.440$ \\
118.0148 & $-47.951 \pm 6.251$ & 125.9233 & $-34.387 \pm 2.657$ & 130.1314 & $-69.869 \pm 2.277$ \\
118.0710 & $-48.042 \pm 3.267$ & 125.9373 & $-34.510 \pm 2.742$ & 130.1447 & $-69.998 \pm 2.608$ \\
118.0877 & $-51.834 \pm 5.119$ & 125.9506 & $-34.692 \pm 2.958$ & 130.1865 & $-69.660 \pm 3.002$ \\
118.1010 & $-53.321 \pm 7.217$ & 126.0339 & $-33.610 \pm 2.408$ & 130.1996 & $-70.063 \pm 2.521$ \\
118.1454 & $-50.003 \pm 2.466$ & 126.0469 & $-33.979 \pm 2.359$ & 130.2126 & $-70.230 \pm 3.305$ \\
118.1585 & $-48.447 \pm 5.900$ & 126.0611 & $-33.347 \pm 2.395$ & 131.9124 & $-61.202 \pm 2.797$ \\
118.1775 & $-53.136 \pm 3.112$ & 126.0743 & $-33.101 \pm 2.326$ & 131.9254 & $-62.224 \pm 3.068$ \\
118.9354 & $-49.826 \pm 3.501$ & 126.1222 & $-32.552 \pm 2.343$ & 131.9384 & $-58.133 \pm 2.653$ \\
118.9478 & $-49.141 \pm 3.659$ & 126.1352 & $-31.563 \pm 2.088$ & 131.9514 & $-59.732 \pm 3.326$ \\
118.9889 & $-50.111 \pm 3.441$ & 126.1482 & $-31.939 \pm 2.511$ & 131.9805 & $-57.723 \pm 3.156$ \\
119.0014 & $-50.441 \pm 3.921$ & 126.1848 & $-31.483 \pm 2.923$ & 132.0066 & $-57.949 \pm 3.367$ \\
119.0138 & $-52.065 \pm 3.781$ & 126.1977 & $-30.767 \pm 2.565$ & 132.0322 & $-56.251 \pm 2.923$ \\
119.0552 & $-51.831 \pm 3.906$ & 126.2107 & $-31.893 \pm 2.842$ & 132.0453 & $-54.676 \pm 2.862$ \\
119.0677 & $-53.398 \pm 3.627$ & 126.9419 & $-36.272 \pm 2.603$ & 132.0583 & $-57.021 \pm 2.694$ \\
119.0802 & $-54.218 \pm 4.755$ & 126.9683 & $-35.469 \pm 2.878$ & 132.0735 & $-55.879 \pm 2.755$ \\
119.1138 & $-55.934 \pm 3.703$ & 127.0924 & $-39.613 \pm 2.182$ & 132.0864 & $-55.474 \pm 2.963$ \\
119.1269 & $-54.718 \pm 3.383$ & 127.1062 & $-39.544 \pm 3.559$ & 132.0995 & $-52.365 \pm 3.715$ \\
119.1394 & $-54.888 \pm 3.520$ & 127.1201 & $-38.081 \pm 2.531$ & 132.1382 & $-53.565 \pm 3.106$ \\
119.9193 & $-84.329 \pm 8.318$ & 127.1465 & $-40.720 \pm 2.797$ & 132.1517 & $-53.728 \pm 3.029$ \\
119.9325 & $-75.408 \pm 3.151$ & 127.1601 & $-38.836 \pm 2.407$ & 132.1713 & $-53.901 \pm 3.810$ \\
120.0306 & $-76.237 \pm 2.404$ & 127.1734 & $-41.524 \pm 2.751$ & 132.2137 & $-49.692 \pm 3.727$ \\
120.0431 & $-75.741 \pm 2.576$ & 127.2103 & $-41.617 \pm 3.177$ & 132.2267 & $-54.489 \pm 3.764$ \\
120.0557 & $-75.138 \pm 2.346$ & 127.2234 & $-41.710 \pm 3.056$ & 132.2397 & $-51.464 \pm 3.900$ \\
120.1009 & $-75.197 \pm 2.393$ & 127.9268 & $-58.126 \pm 4.144$ & 132.9132 & $-59.635 \pm 3.026$ \\
120.1135 & $-74.375 \pm 2.652$ & 127.9398 & $-59.568 \pm 3.727$ & 132.9274 & $-57.425 \pm 3.343$ \\
120.1260 & $-75.235 \pm 2.312$ & 127.9528 & $-59.855 \pm 4.283$ & 132.9404 & $-56.218 \pm 2.664$ \\
120.1824 & $-75.016 \pm 2.495$ & 128.0551 & $-62.303 \pm 4.473$ & 133.0305 & $-54.996 \pm 2.741$ \\
120.1962 & $-74.390 \pm 2.519$ & 128.0698 & $-61.858 \pm 4.231$ & 133.0435 & $-53.878 \pm 2.518$ \\
120.2103 & $-74.013 \pm 2.769$ & 128.0829 & $-61.410 \pm 4.222$ & 133.0570 & $-54.587 \pm 2.539$ \\
120.2417 & $-72.921 \pm 2.729$ & 128.1450 & $-64.184 \pm 4.351$ & 133.9029 & $-35.155 \pm 3.759$ \\
120.2543 & $-72.066 \pm 2.847$ & 128.1585 & $-61.980 \pm 3.912$ & 133.9158 & $-36.872 \pm 3.280$ \\
120.2669 & $-74.260 \pm 3.256$ & 128.1715 & $-60.965 \pm 4.288$ & 133.9288 & $-37.601 \pm 3.099$ \\
120.9177 & $-69.011 \pm 2.892$ & 128.1846 & $-65.137 \pm 3.990$ & 133.9773 & $-36.505 \pm 3.376$ \\
120.9298 & $-67.862 \pm 2.769$ & 128.2014 & $-63.062 \pm 3.888$ & 133.9903 & $-35.226 \pm 3.121$ \\
120.9430 & $-68.414 \pm 2.985$ & 128.2145 & $-63.871 \pm 4.560$ & 134.0064 & $-36.135 \pm 3.172$ \\
120.9659 & $-67.933 \pm 2.808$ & 128.2275 & $-63.709 \pm 4.447$ & 134.0861 & $-37.570 \pm 3.744$ \\
120.9789 & $-68.716 \pm 2.809$ & 128.9267 & $-69.557 \pm 3.368$ & 134.9085 & $-28.154 \pm 3.774$ \\
120.9929 & $-67.612 \pm 2.848$ & 128.9396 & $-69.768 \pm 3.061$ & 134.9215 & $-29.949 \pm 2.582$ \\
121.0343 & $-67.387 \pm 2.906$ & 128.9526 & $-70.041 \pm 3.134$ & 134.9346 & $-29.342 \pm 2.227$ \\
122.1584 & $-68.859 \pm 5.309$ & 129.0007 & $-68.518 \pm 3.256$ & 134.9852 & $-28.680 \pm 2.028$ \\
122.1803 & $-76.832 \pm 5.303$ & 129.0142 & $-70.118 \pm 3.534$ & 134.9982 & $-29.925 \pm 2.816$ \\
122.1934 & $-70.372 \pm 3.793$ & 129.0272 & $-68.765 \pm 3.262$ & 135.0146 & $-29.764 \pm 2.670$ \\
123.9486 & $-55.418 \pm 2.464$ & 129.1112 & $-68.900 \pm 2.794$ & 135.0596 & $-30.751 \pm 3.978$ \\
123.9680 & $-53.818 \pm 2.188$ & 129.1244 & $-68.574 \pm 2.835$ &   &  \\ \hline
\end{tabular}
\end{table*}

\end{document}